\documentclass[doublecol]{epl2}

\usepackage{mathtools}
\usepackage{amssymb}
\usepackage{amsfonts}
\usepackage{fancyhdr}
\usepackage{slashed}
\usepackage{bbm}
\usepackage{latexsym,epsfig,bbm}
\usepackage{amsthm}
\usepackage{braket}

\usepackage[colorlinks=true,urlcolor=blue,citecolor=blue]{hyperref}
\newcommand\DmQn{\mbox{\textit{DmQn}}}

\DeclareRobustCommand{\rchi}{{\mathpalette\irchi\relax}}
\newcommand{\irchi}[2]{\raisebox{\depth}{$#1\chi$}} 

\title{ \vspace{-2em}
Decomposition of Nonlinear Collision Operator in Quantum \\ Lattice Boltzmann Algorithm
}
\shorttitle{QLB Algorithm with Nonlinear Collision} 

\author{E. Dinesh Kumar\inst{*} \and Steven H. Frankel\inst{\dagger}}
\shortauthor{Kumar and Frankel}

\institute{                    
  Faculty of Mechanical Engineering \\ Technion - Israel Institute of Technology, Haifa, 3200003, Israel.
  \\
  $^*$\email{edk261@gmail.com},  $^\dagger$\email{frankel@technion.ac.il} 
  \vspace{-2em}
}

\abstract{ 
We propose a quantum algorithm to tackle the quadratic nonlinearity in the Lattice Boltzmann (LB) collision operator. The key idea is to build the quantum gates based on the particle distribution functions (PDF) 
within the coherence time for qubits. Thus, both the operator and a state vector are linear functions of PDFs, and upon quantum state evolution, the resulting PDFs will have quadraticity. 
To this end, we decompose the collision operator for a $DmQn$ lattice model into a product of $2(n+1)$ operators, where $n$ is the number of lattice velocity directions. 
After decomposition, the $(n+1)$ operators with constant entries remain unchanged throughout the simulation, whereas the remaining $(n+1)$ will be built based on the statevector of the previous time step.
Also, we show that such a decomposition is not unique.  Compared to the second-order Carleman-linearized LB, the present approach reduces the circuit width by half and circuit depth by exponential order. 
The proposed algorithm has been verified through the one-dimensional flow discontinuity and two-dimensional Kolmogrov-like flow test cases.
}

\begin{document}
\maketitle

\section{Introduction}

The potential for simulating fluid flows on quantum computers has generated significant interest due to the unique properties of qubits, such as superposition and entanglement. By encoding flow variables onto the probability amplitudes of qubit states, we can achieve significantly reduced memory requirements and a potentially exponential increase in computing speed. However, quantum operations are inherently linear and unitary, while the equations of fluid mechanics exhibit nonlinearity. As a result,  existing computational fluid dynamics (CFD) algorithms need to be redesigned and adapted to fully utilize the power of quantum processor units (QPUs).
\par
Traditional CFD methods solve the discretized Navier-Stokes equation (NSE), and commonly used algorithms are implicit in time-stepping, which requires linear system solvers (LSS)
\cite{HHL2009, Cao2012, bharadwaj2023PNAS, ingelmann2024, bharadwaj2024compact}. The Lattice Boltzmann Method (LBM) has been considered an alternate approach to NSE-based CFD solvers and has been studied extensively for various flow problems. Due to locality in space and explicit time-stepping, LBM does not require LSS and has proven to be efficient on graphic processor units (GPUs) \cite{kuznik2010lbm}.
Compared to NSE, where the nonlinear term $(\mathbf{u}\cdot \nabla{\mathbf{u}}$) is also non-local, the LB equation has nonlinearity ($\mathbf{u}\cdot\mathbf{u}$) in the equilibrium function and is local. This nonlinearity does not pose any difficulty in classical computers, as  $\mathbf{u}$ will be saved as a floating point number followed by a square operation \cite{Steijl2022, steijl2024floating}. However, if we mimic the classical procedure on QPUs, the quantum advantage may be lost, especially when amplitude encoding is chosen. 
\par
There are several existing works on Quantum Lattice Boltzmann (QLB) for fluid flow simulation, but most of them ignore nonlinearity and are only used for very low Reynolds number cases \cite{ Mezzacapo2015, budinski2021ADE, budinski2022NSE, Schalkers2024JCP, dinesh2024_linear}. In the absence of nonlinearity, the particle distribution functions (PDF) $\boldsymbol{f}$ can be encoded directly as probability amplitudes of qubits. 
However, evaluating the nonlinear term involves the vector $\boldsymbol{f} \otimes \boldsymbol{f}$. \cite{Succi2023} highlighted that it is impossible to have a linear operator that performs the squares of probability amplitudes, as the resulting state vector may not be a unit vector. Furthermore, the no-cloning theorem states that it is impossible to copy an arbitrary quantum state to create an identical quantum state \cite{Succi2023}.
To address this, the Carleman linearization (CL) technique can be utilized which converts the finite-dimensional nonlinear system into a infinite-dimensional linear system which is then truncated for implementation \cite{carleman1932}. Previously, CL has been successfully applied to LBM up to second-order truncation for low to moderate Reynolds number cases. When CL is applied to LBM, the resultant state vector becomes infinite, i.e., $(\boldsymbol{f}, \boldsymbol{f}  \otimes \boldsymbol{f}, \boldsymbol{f}  \otimes \boldsymbol{f}  \otimes \boldsymbol{f}, \ldots)$. 
In addition, another approximation has to be made for the term $\frac{1}{\rho} \approx 2-\rho$, where $\rho$ is the density, which is valid for weakly incompressible regimes. Even after successive simplifications, the Carleman LB (CLB) algorithm still requires repeated encoding and read-out of state vectors for each time step \cite{itani2022, itani2024, Claudio2024POF, Claudio2024}.
\par
In the present work, instead of the CL algorithm, we propose the decomposition of the LB operator as a product of multiple operators derived from the PDFs from the previous time step. We retain the first-order Taylor approximation for $\frac{1}{\rho}$ and accept that the encoding and read-out process is required for each time step \cite{itani2024, Claudio2024POF, Claudio2024}. Our approach does not involve any truncation of infinite sets. Instead, the state vector is a finite set, specifically, four sets of $\boldsymbol{f}$ appended sequentially. 
\par
In the following, we briefly describe the LB equations, followed by deriving the quadratic formulation of the LB collision operation. Later, we will present the quantum algorithm and the computational complexity of two-qubit gates.  

\section{Lattice Boltzmann Method}
\label{sec_lbm}
LBM describes the evolution of flow field through PDFs in a uniform Cartesian grid with the chosen lattice model $\DmQn$, where $m$ and $n$ denotes the number of spatial dimensions and velocity directions, respectively. The single relaxation time LB model for the flow field is given by
\begin{equation}
	f_i(\mathbf{x} + \mathbf{e}_{i} \Delta t, t+\Delta t)  
	= 
	f_i (\mathbf{x},t) - \frac{\Delta t}{\tau}
	\left[ 
	f_i(\mathbf{x},t) - f_i^{eq}(\mathbf{x},t)
	\right]
	\label{eq_lbm_bgk}
\end{equation}
where $f_i$ represents the PDF along the $i^{th}$ direction, 
$\mathbf{e}_i$ is the lattice velocity, and 
the kinematic viscosity $\nu$ is related to the relaxation parameter  
$\tau = 3\nu + 0.5$. The equilibrium distribution function (EDF) is given by
\begin{equation}
	f_i^{eq} = w_{i} \rho \left[ 1+\frac{\mathbf{e}_i\cdot \mathbf{u}}{c_s^2} + \frac{(\mathbf{e}_i\cdot \mathbf{u})^2}{2c_s^4} - \frac{\mathbf{u}\cdot \mathbf{u}}{2c_s^2} \right]
	\label{eq_feq_ns}
\end{equation}
where $w_i$ is the lattice weight, $c_s$ the sound speed, $\rho$ the fluid density, and $\mathbf{u}$ is the flow velocity. 
The zeroth and first moment of PDFs yields the macroscopic flow variables density($\rho$) and momentum ($\rho\mathbf{u}$),
\begin{align}
		\rho(\mathbf{x}, t) &= \sum_i f_i(\mathbf{x}, t)
		\label{eq_lb_density} \\ 
\rho(\mathbf{x}, t)\mathbf{u}(\mathbf{x}, t) &= \sum_i  \mathbf{e}_i f_i(\mathbf{x}, t)
		\label{eq_lb_momentum}
\end{align}
Typically, LB algorithm is split into collision and streaming steps. In the collision step, 
the interaction of particles and their relaxation to equilibrium state is modelled. 
Later the particles move to their neighbouring site according to the lattice velocity, a process termed streaming.
Thus, Eq.~\eqref{eq_lbm_bgk} can be written as 
\begin{align}
	f_i^*(\mathbf{x}, t) &= 
	f_i (\mathbf{x},t) - \frac{\Delta t}{\tau} \left[ f_i(\mathbf{x},t) - f_i^{eq}(\mathbf{x},t) \right]
	\label{eq_lb_coll}
\\
f_i(\mathbf{x}, t+\Delta t)  &= f_i^*(\mathbf{x} - \mathbf{e}_{i} \Delta t, t) 
\label{eq_lb_strm}
\end{align}
where $f_i^*$ is the post-collision state. 
In the presence of a wall boundary, we use half-way bounce back scheme to reconstruct PDFs coming from the wall node ($f_{-i}$), 
\begin{align}
f_{-i}(\mathbf{x}_b, t+\Delta t)  &=  f^*_i(\mathbf{x}, t)
\label{eq_mov_wall}
\end{align}

\section{Formulation}
\label{sec_lbm}
\subsection{Quadratic form for collision step}
\label{sec_quadratic}

First, we rewrite Eq.~\eqref{eq_feq_ns} and 
using Eqs.~\eqref{eq_lb_density}~and~\eqref{eq_lb_momentum}, we get the quadratic form for EDF as,
\begin{align}
	f_i^{eq} &= \frac{1}{\rho} w_{i} \left[ \rho^2 + \rho \frac{\mathbf{e}_i\cdot \rho \mathbf{u}}{c_s^2} + \frac{(\mathbf{e}_i\cdot \rho \mathbf{u})^2}{2c_s^4} - \frac{\rho \mathbf{u}\cdot \rho \mathbf{u}}{2c_s^2} \right] \\
	&= \frac{1}{\rho} \sum_j \biggl( \sum_{\substack{k \\ j \le k}} \alpha_{ijk} f_k  \biggl) f_j 
	\label{eq_feq_ns_facr}
\end{align}
where the coefficient $\alpha_{ijk}$ is given by
\begin{multline}
\alpha_{ijk} = w_{i} \left(\frac{1}{2} \right)^{\delta_{jk}} \left[ 2 + \frac{1}{c_s^2} \left( \mathbf{e}_i\cdot\mathbf{e}_j  + \mathbf{e}_i\cdot\mathbf{e}_k  - \mathbf{e}_j\cdot\mathbf{e}_k  \right)
     \right.\\ \left.\vphantom{\int}
  +  \frac{1}{c_s^4} (\mathbf{e}_i\cdot\mathbf{e}_j)(\mathbf{e}_i\cdot\mathbf{e}_k) \right]
\label{eq_coeff_alpha}
\end{multline}
and $\delta_{jk}$ denotes the Kronecker delta function. Next, the expression for collision in Eq.~\eqref{eq_lb_coll} can be rewritten as, 
\begin{align}
	f_i^* &= \frac{1}{\rho} \left( 1- \frac{{\Delta t}}{\tau} \right) \rho f_i + \frac{{\Delta t}}{\tau} f_i^{eq} 
\label{eq_coll_exp_0}
\end{align}
Using Eqs.~\eqref{eq_lb_density}~and~\eqref{eq_coeff_alpha}, the quadratic form of the collision step can be written as 
\begin{align}
f_i^* &= \frac{1}{\rho} 
        {
        \begin{pmatrix}
        f_1 \\
        f_2 \\
        \vdots \\
        f_{n_{\mathbf{e}}}
        \end{pmatrix}
        }^{\mathbf{T}}
        \begin{pmatrix}
        \beta_{i11} & \beta_{i12} & \ldots & \beta_{i1n_{\mathbf{e}}} \\
                    & \beta_{i22} & \ldots & \beta_{i2n_{\mathbf{e}}} \\
       \multicolumn{2}{c}{ \mbox{\Large\( 0 \)}   }       & \ddots & \vdots                  \\  
                   &            &        & \beta_{in_{\mathbf{e}}n_{\mathbf{e}}}
        \end{pmatrix}
        {
        \begin{pmatrix}
        f_1 \\
        f_2 \\
        \vdots \\
        f_{n_{\mathbf{e}}}
        \end{pmatrix}
        }
\label{eq_coll_quad}
\end{align}
where $\beta_{ijk} = {\gamma_{_{ijk}}} \cdot \left( 1- \frac{{\Delta t}}{\tau} \right) + \rchi_{jk} \cdot \frac{{\Delta t}}{\tau} \alpha_{ijk}$ and $\mathbf{T}$ denotes the transpose. The indicator functions $\gamma$ and $\rchi$ for  $i^{\text {th}}$ lattice direction  is given by,
\begin{align}
 \gamma_{_{ijk}} &= \delta_{ij} + \delta_{ik} - \delta_{ij}\cdot\delta_{ik}
\\
	\rchi_{jk} &=
	\begin{cases}
		1 & \text{if } j \leq k \\
		0 & \text{otherwise } 
	\end{cases} 
\end{align}

\subsection{Matrix formulation for weakly compressible LBM}
\label{subsec_matvec_lbm}
In case of weakly compressible flows, $\rho^{-1} \approx 2-\rho$, we then express
\begin{align}
(2-\rho) &= (-1,-1,\cdots,-1,2) \cdot (f_1,f_2,\cdots,f_{n_{\mathbf{e}}},1)^{\mathbf{T}}
\label{eq_mtrx_2rho}
\end{align}  
where the value $1$ appended at the end of PDFs can be treated as a auxiliary constant.
Therefore, the collision step can be written as
\begin{align}
\begin{pmatrix}
f^*_i \\
1
\end{pmatrix}
&= 
{
\begin{pmatrix}
\boldsymbol{f} \\
1
\end{pmatrix}
}^\mathbf{T}
\begin{pmatrix}
\boldsymbol{\beta}_i & 0 \\
0 & 1
\end{pmatrix}
\begin{pmatrix}
\mathsf{Diag}(\boldsymbol{f}) & 0 \\
0 & 1
\end{pmatrix}
\mathbf{W}
{
\begin{pmatrix}
\boldsymbol{f} \\
1
\end{pmatrix}
}
\label{eq_coll_quad_fin}
\end{align}
where $\boldsymbol{f}=(f_1,f_2,\cdots,f_{n_{\mathbf{e}}})$, and 
$\boldsymbol{\beta}_i$ represent the upper triangular matrix in Eq.~\eqref{eq_coll_quad}.
The term $\mathsf{Diag}(\boldsymbol{f})$ represents the diagonal matrix with entries from ${\boldsymbol{f}}$. 
The matrix $\mathbf{W}$ is given by
\begin{align}
\mathbf{W} &=
\begin{pmatrix}
-1 & -1 & \cdots & -1 & 2 \\
\vdots & \vdots & & \vdots & \vdots  \\
-1 & -1 &\cdots & -1 & 2 \\
0 & 0 &\cdots & 0 & 1 \\
\end{pmatrix}
\end{align}

\section{Quantum Algorithm}
\label{sec_quan_algo}
\subsection{Encoding}
\label{subsec_encode}
Let $n_g$ denote the total number of grid points and the number of PDFs will be $n_f = n_{\mathbf{e}} n_g$.
According to the lattice directions, we arrange and order the PDFs in a vector form,
$\boldsymbol{df} = (\boldsymbol{f}_1, \boldsymbol{f}_2, \ldots, \boldsymbol{f}_{n_{\mathbf{e}}},1)$, where the suffix refers to the direction, and each $\boldsymbol{f}_i$ is of length $n_g$. Since PDFs are encoded as a quantum state-vector, we define 
$\phi = (\boldsymbol{df}, \boldsymbol{df}, \boldsymbol{df}, \boldsymbol{df})$.
A computational register containing 
$n_q = \log_2(n_f)$ qubits along with two ancilla qubits will be required to encode the PDFs. 
The initial state to encode will be 
\begin{equation}
	\ket{\boldsymbol{\phi}^0} = \ket{0}_a \ket{0}_a \sum_{i=1}^{2^{n_q}} \frac{\phi_i}{\lVert \phi_i \rVert} \ket{i}_q
\end{equation}

%
\subsection{Collision}

Before we express matrix version of Eq.~\eqref{eq_coll_quad_fin}, we define the following matrices:
\begin{align}
\widetilde{\mathbf{W}} &=
\begin{pmatrix}
-\mathbf{I}^{^g}_{11} & -\mathbf{I}^{^g}_{12} & \cdots & -\mathbf{I}^{^g}_{1{n_{\mathbf{e}}}} & \mathbf{2}^{^g}_{1} \\
-\mathbf{I}^{^g}_{21} & -\mathbf{I}^{^g}_{22} & \cdots & -\mathbf{I}^{^g}_{2{n_{\mathbf{e}}}} & \mathbf{2}^{^g}_{2} \\
\vdots & \vdots & & \vdots & \vdots  \\
-\mathbf{I}^{^g}_{{n_{\mathbf{e}}}1} & -\mathbf{I}^{^g}_{{n_{\mathbf{e}}}2} & \cdots & -\mathbf{I}^{^g}_{{n_{\mathbf{e}}}{n_{\mathbf{e}}}} &  \mathbf{2}^{^g}_{n_{\mathbf{e}}} \\
0 & 0 &\cdots & 0 & 1 \\
\end{pmatrix}
\\
\widetilde{\mathbf{D}} &= \mathsf{Diag}(\boldsymbol{df})
\end{align}
\begin{align}
\widetilde{\mathbf{B}}_i &=
\begin{pmatrix}
\tilde{\boldsymbol{\beta}}_{1i1} & 	\tilde{\boldsymbol{\beta}}_{1i2} & \cdots & \tilde{\boldsymbol{\beta}}_{1i{n_{\mathbf{e}}}} & \mathbf{0}^{^g}_{1} \\ 
\tilde{\boldsymbol{\beta}}_{2i1} & \tilde{\boldsymbol{\beta}}_{2i2} & \cdots & \tilde{\boldsymbol{\beta}}_{2i{n_{\mathbf{e}}}} & \mathbf{0}^{^g}_{2} \\ 
\vdots & \vdots & & \vdots & \vdots \\
\tilde{\boldsymbol{\beta}}_{n_{\mathbf{e}}i1} & \tilde{\boldsymbol{\beta}}_{n_{\mathbf{e}}i2} & \cdots & \tilde{\boldsymbol{\beta}}_{{n_{\mathbf{e}}}i{n_{\mathbf{e}}}} & \mathbf{0}^{^g}_{n_{\mathbf{e}}} \\
0 & 0 &\cdots & 0 & 1 
\end{pmatrix}
\\
\widetilde{\mathbf{D}}_i &= \mathsf{Diag}(\boldsymbol{f}_i)
\end{align}
\begin{align}
\widetilde{\mathbf{F}}_i &=
\begin{pmatrix}
(\widetilde{\mathbf{D}}_i)_{11} &  &  &  &\\
 & (\widetilde{\mathbf{D}}_i)_{22} &  &  &  & \\
 & & \ddots &  & \\
 &  & & (\widetilde{\mathbf{D}}_i)_{{n_{\mathbf{e}}}{n_{\mathbf{e}}}} & \\
 &  &  &  & \delta_{i1} \\
\end{pmatrix}
\end{align}
\noindent
where $\mathbf{I}^{^g}_{rc}$ denotes the identity matrix of size $n_g \times n_g$, 
$\mathbf{2}^{^g}_{i} = (2, \cdots, 2)$ and $\mathbf{0}^{^g}_{i} = (0, \cdots, 0)$ are the vectors of size $n_g$.
The term $\tilde{\boldsymbol{\beta}}_{ric} = \rchi_{ic}\beta_{ric}\mathbf{I}^{^g}_{rc}$, where 
$\beta_{ric}$ is defined in Eq.~\eqref{eq_coll_quad}, with $r$ and $c$ represent row and column 
indices, respectively. 
Based on the above set of matrices, we further define 
\begin{align}
\begin{split}
    \widehat{\mathbf{W}} &= 
    \begin{pmatrix}
    \widetilde{\mathbf{W}} & & &  \\
    & \widetilde{\mathbf{W}} & & \\
    & & \widetilde{\mathbf{W}} & \\
    & & & \widetilde{\mathbf{W}} \\
    \end{pmatrix} ;
    \\
    \widehat{\mathbf{B}}_1 &= 
    \begin{pmatrix}
    \widetilde{\mathbf{B}}_1 &  &  &  \\
    & \mathbf{I} &  &  \\
    & & \mathbf{I} & \\
    & & & \mathbf{I} \\
    \end{pmatrix} ;
    \\
    \widehat{\mathbf{B}}_{i} &= 
    \begin{pmatrix}
    \mathbf{I} & \mathbf{0} & \mathbf{0} & \mathbf{0} \\
    & \mathbf{0} & \widetilde{\mathbf{B}}_i & \mathbf{0} \\
    & & \mathbf{I} & \mathbf{0}\\
    & & & \mathbf{I} \\
    \end{pmatrix} ;
\end{split}
\begin{split} 
    \widehat{\mathbf{D}} &= 
    \begin{pmatrix}
    \widetilde{\mathbf{D}} & & & \\
    & \widetilde{\mathbf{D}} & & \\
    & & \widetilde{\mathbf{D}} & \\
    & & & \widetilde{\mathbf{D}} \\
    \end{pmatrix}
    \\
    \widehat{\mathbf{F}}_1 &= 
    \begin{pmatrix}
    \widetilde{\mathbf{F}}_1 &  &  &   \\
    & \mathbf{I} &  &  \\
    & & \mathbf{I} & \\
    & & & \mathbf{I} \\
    \end{pmatrix}
    \\
    \widehat{\mathbf{F}}_{i} &= 
    \begin{pmatrix}
    \mathbf{I} & \widetilde{\mathbf{F}}_i & \mathbf{0} & \mathbf{0} \\
    & \mathbf{0} & \mathbf{I} & \mathbf{0} \\
    & & \mathbf{I} & \mathbf{0}\\
    & & & \mathbf{I} \\
    \end{pmatrix}
\end{split}
\label{eq_hat_mat}
\end{align}
Therefore, the collision operator is given by 
\begin{align}
\ket{\boldsymbol{\phi}^*} &= \widehat{\mathbf{F}}_{n_{\mathbf{e}}} \text{ } \widehat{\mathbf{B}}_{n_{\mathbf{e}}} \cdots \text{ } 
            \widehat{\mathbf{F}}_{2} \text{ } \widehat{\mathbf{B}}_{2} \text{ }  \widehat{\mathbf{F}}_{1} \text{ } \widehat{\mathbf{B}}_{1}\text{ } 
            \widehat{\mathbf{D}} \text{ } \widehat{\mathbf{W}} \text{ } \ket{\boldsymbol{\phi}^0}  
\label{eq_decom_coll}
\end{align}
In Eq.~\eqref{eq_decom_coll}, when the SV is multiplied by $\widehat{\mathbf{W}}$ yields the value $2-\rho$, which then multiplied by $\widehat{\mathbf{D}}$ results in the first order approximation of $\frac{f_i}{\rho}$, i.e., the column vector defined on the right side of 
Eq.~\eqref{eq_coll_quad}. Upon successive multiplication by $\widehat{\mathbf{F}}_{i} \text{ } \widehat{\mathbf{B}}_{i}$, yields the PDFs in the $i^{\text{th}}$ direction. Thus, for $n_{\mathbf{e}}$ velocity directions, the collision operation is 
decomposed into $2n_{\mathbf{e}}+2$ operations. It is important to note that the state vector has been initialized with four sets of $\boldsymbol{df}$. Among these, one will ultimately lead to the final outcome, while another is designated for storing temporary values. The remaining two sets remain constant, providing the initial state vector values during intermediate computations. Consequently, the matrices defined in Eq.~\eqref{eq_hat_mat} are not unique.  In particular, the following is a valid choice:
\begin{align}
\begin{split}
    \widehat{\mathbf{B}}_{i} &= 
    \begin{pmatrix}
    \mathbf{I} & \mathbf{0} & \mathbf{0} & \mathbf{0} \\
    & \mathbf{0} & \mathbf{I} & \mathbf{0} \\
    & \widetilde{\mathbf{B}}_i & \mathbf{0} & \mathbf{0}\\
    & & & \mathbf{I} \\
    \end{pmatrix}
\end{split}
;
\begin{split}
    \widehat{\mathbf{F}}_{i} &= 
    \begin{pmatrix}
    \mathbf{I} &  \mathbf{0} & \widetilde{\mathbf{F}}_i & \mathbf{0} \\
    & \mathbf{0} & \mathbf{0} & \mathbf{I} \\
    & & \mathbf{I} & \mathbf{0}\\
    & & & \mathbf{I} \\
    \end{pmatrix}
\end{split}
\end{align}
\subsection{Streaming and Boundary Conditions}
The streaming matrix $S$ consists of $n_{\mathbf{e}} \times n_{\mathbf{e}}$
blocks with the size of each block $n_g \times n_g$. The matrix $S$ is square binary matrix 
where each row/column contains one entry equal to 1. Since every row index $i$ can be decomposed as 
$x + y n_x + i_e n_x n_y$, and the corresponding column index $j$ such that $s_{ij}=1$, can be computed by replacing the co-ordinate $\mathbf{x}$ by $\mathbf{x} - \mathbf{e}_{i} \Delta t$. In the present study, we utilized the 
wall boundary condition, accordingly the column index will be computed as 
$x_{\Delta} + y_{\Delta} n_x + i_{-e} n_x n_y$, where 
$(x_{\Delta}, y_{\Delta}) = \mathbf{x} - \mathbf{e}_{i}$ and $i_{-e}$ is the reflection of the direction $i_e$. 
Based on the matrix $S$, we define the streaming operator as
\begin{align}
	\begin{split}
    \widehat{\mathbf{S}} &= 
	\begin{pmatrix}
		\widetilde{\mathbf{S}} &  &  &  \\
		& \mathbf{I} &  &  \\
		& & \mathbf{I} & \\
		& & & \mathbf{I} \\
	\end{pmatrix}
\end{split}
\begin{split}
\text{ ; where  }
		\widetilde{\mathbf{S}} &= 
			\begin{pmatrix}
				S &  \\
				  & 1 \\
			\end{pmatrix}
\end{split}
\label{eq_decom_strm}
\end{align}
From Eq.~\eqref{eq_decom_strm}, it is clear that streaming is performed only on the computational qubits without any involvement from ancilla qubits.
Thus, the PDF for the next time step can be obtained from the quantum state,
\begin{align}
 	\ket{\boldsymbol{\phi}^n} &=  \widehat{\mathbf{S}} \ket{\boldsymbol{\phi}^*} 
 	\label{eq_q_stream}
\end{align}

\section{Results and Discussion}
\subsection{Verification - \noindent Case 1: 1D discontinuity flow}
\begin{figure}[h!]
	\centering
	\includegraphics[width=0.49\textwidth]{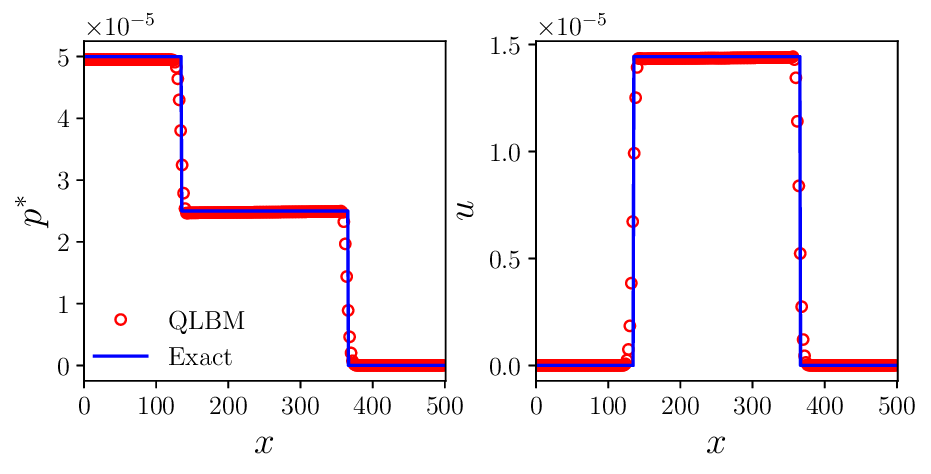}
	\caption{Comparison of normalized pressure and velocity of 1D discontinuity flow obtained from QLB with the exact Riemann solution at $200^{\text{th}}$ time step.}
	\label{fig_shock}
\end{figure}

The classical shock tube discontinuity problem in one dimension is used to verify the present the algorithm. Because of the weak compressibility assumption, we refer it as a discontinuity rather than a shock. The one dimensional domain of size $n_g=500$ grid points and the $D1Q3$ lattice has been chosen. At the initial time, the velocity is set to zero everywhere and density is defined as, 
\begin{align}
\rho (x,0) &=
    \begin{cases}
        1.0 + \Delta\rho & \text{if } x \leq \frac{n_g}{2} \\
        1.0 & \text{otherwise } 
    \end{cases} 
\end{align}
where $\Delta\rho=5\times10^{-5}$ is chosen to satisfy the weak incompressibility assumption made earlier. 
Bounce-back boundary condition is applied at the both ends of the domain. After simulating
$200$ time steps, we obtained a good comparison with the exact Riemann solution for normalized pressure ($p^* = (p-p_s)/p_s$; $p_s=\Delta\rho {c_s^2}$) 
and velocity ($u$) \cite{Toro2013} and is given in Fig.~\ref{fig_shock}. The relative difference between exact and QLB results is to found to be 5\%. 
\subsection{Case 2: 2D Kolmogorov flow}
\begin{figure}[h!]
	\centering
	\includegraphics[scale=0.4]{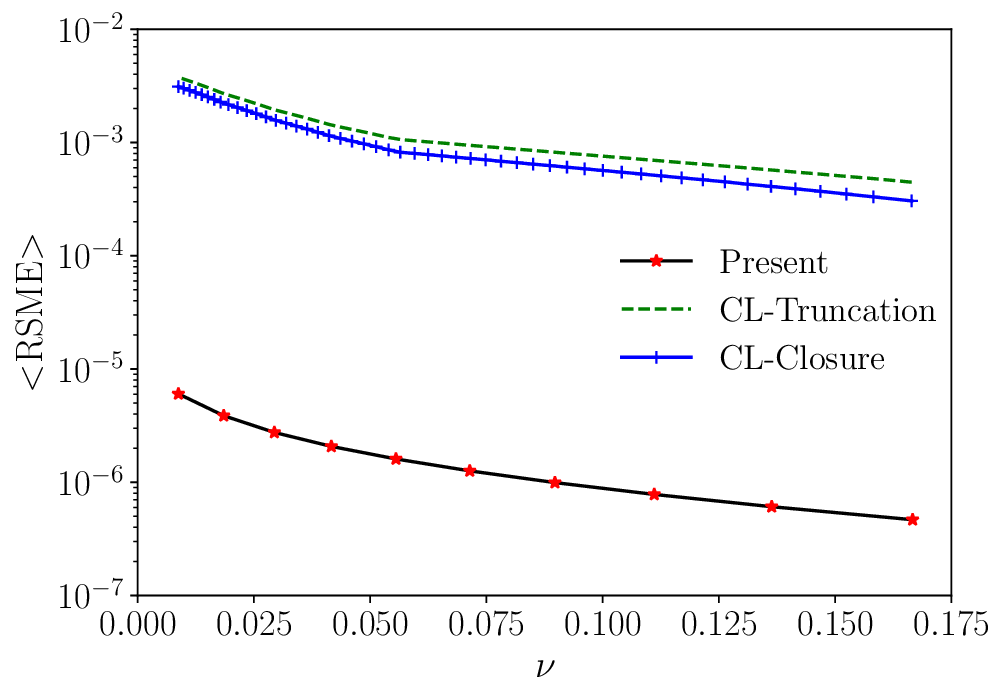}
	\caption{Comparison of root mean square error obtained for various viscosity values are compared with thre results of carleman second order truncation and closure approaches presented in \cite{Claudio2024}.}
	\label{fig_kolmogrov}
\end{figure}

In order to compare the accuracy of present algorithm with that of Carleman linearized - LBM presented in 
\cite{Claudio2024}, we performed the two-dimensional simulation of Kolmogorov-like flow on a $32\times32$ grid. The D2Q9 lattice is
used.
Initially, the PDFs are defined as,
\begin{multline}
f_i(x,y) = 
w_i \left[ 
        1 + A_x \cos\left( \frac{2\pi k_x}{N_y} y \right) \mathbf{e}_i \cdot \mathbf{e}_x +
             \right.\\ \left.\vphantom{\int}
            A_y \cos\left( \frac{2\pi k_y}{N_x} x \right) \mathbf{e}_i \cdot \mathbf{e}_y
    \right]
\label{eq_kolmogrov}
\end{multline}
where $\mathbf{e}_x=(1,0)$ and $\mathbf{e}_y=(0,1)$. The parameters in Eq.~\eqref{eq_kolmogrov} are taken as: $A_x=0.3$, $A_y=0.2$, $k_x=1$, and $k_y=4$.
We choose different viscosity values $\nu$ ranging from $1/6$ to $0.0088$, and for each $\nu$ we ran the simulation up to 100 time steps.  
After simulating 100 time steps, we compute the mean value of Root Mean Squared Error (RMSE) between the distribution functions obtained from the present algorithm ($f^{p}_i$) and the exact LBM simulation ($f^{e}_i$),
\begin{align}
	<\text{RSME}> = \sum_{i=1}^{n_{\mathbf{e}}} \frac{1}{n_{\mathbf{e}}} 
					\sqrt{\sum_{j=1}^{n_g} \frac{1}{n_g} 
							\left( \frac{  {f^{p}_{i_j} }- {f^{e}_{i_j} } } { {f^{e}_{i_j} } }  \right) }
\end{align} 
Since CL transforms a finite set to an infinite dimensional set of equations, \cite{Claudio2024} 
analysed two different approaches, truncation (neglecting higher order terms) and 
closure (approximate the product of functions into products of a function and a constant). Fig.~\ref{fig_kolmogrov} shows the comparison of RSME with the 
results of Carleman second order truncation and closure approaches presented in \cite{Claudio2024}. Since the present approach does not require any order of truncation
except the assumption on weak compressiblity, the $<\text{RSME}>$ value is below $10^{-5}$ compared to $10^{-2}$ from Carleman closure approach. 

\subsection{Computational Complexity}

\begin{figure}[h!]
	\centering
	\includegraphics[scale=0.55]{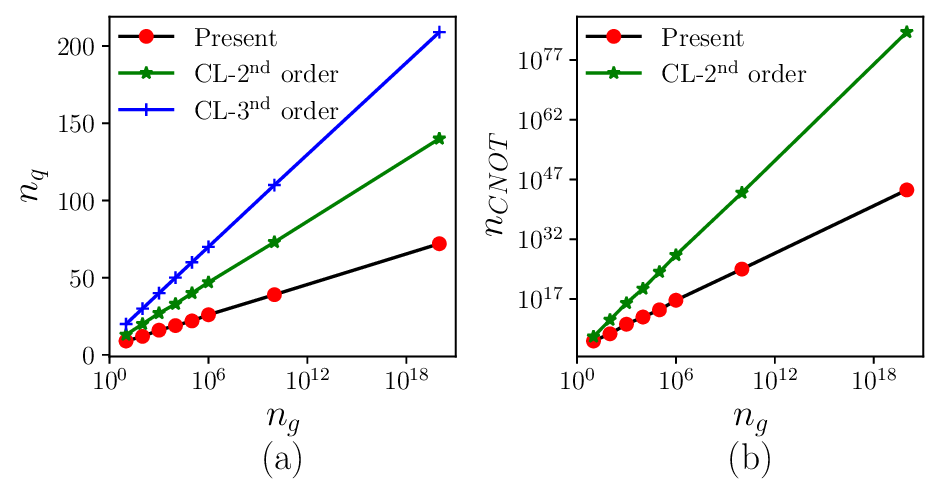}
	\caption{Comparison of (a) number of qubits and (b) number of CNOT gates required for number of grid points between the present algorithm \textit{vs.} the Carleman second and third order truncation.}
	\label{fig_compare_carleman}
\end{figure}

In the present approach, the number of qubits required will be $2+\log_2(n_f)$, whereas the Carleman truncated system of $k^{\text{th}}$ order would require $1+\log_2(n_f + n_f^2 + \ldots + n_f^k)$ qubits. Fig.~\ref{fig_compare_carleman}a shows the number of qubits required for the grid size ranging from $10^1$ to $10^{20}$ using D2Q9 lattice. For larger grid sizes, the CL- second (CL2) and third (CL3) order truncation approach requires almost twice and thrice the qubit resource than the present approach.  
\par
The number of two-qubit gates is one of the ways to estimate the complexity of the present quantum algorithm. Since PDFs are encoded as amplitudes of the qubit states, the state preparation step is required before performing LB operations. According to \cite{shende2005}, $O(2^{n_q})$ CNOT gates are required for state preparation. 
Based on Eq.~\eqref{eq_decom_coll} and \eqref{eq_q_stream}, the number of LB operators per time step is $2n_{\mathbf{e}}+3$, and each operator would require $O(2^{n_q-1}(2^{n_q}-1))$ CNOT gates. In the second-order Carleman truncation, the qubit requirement will be $n^c_q = 1+\log_2(n_f + n_f^2)$. Correspondingly, the $O(2^{n^c_q-1}(2^{n^c_q}-1))$ CNOT gates will be required. Fig.~\ref{fig_compare_carleman}b shows the number of CNOT gates required for different grid sizes. For smaller grids of size $10^3$, the gate counts of CL2 are a few orders higher than the present approach; however, for the larger grid, it is exponentially larger.

\section{Summary}
We have developed a quantum algorithm to deal with the non-linearity in the Lattice Boltzmann collision operator. By decomposing the collision operator, we significantly reduced the circuit width by half and circuit depth in exponential order when compared to the Carleman linearization technique. While there are still challenges to be addressed, such as the encoding and read-out process for each time step and the construction of quantum gates, this work aims to minimize the quantum resources required by the CL technique and prevent the formation of an infinite system. Due to the weakly compressible assumption and the first-order approximation for $1/\rho$, this work is limited to moderate Reynolds numbers of O(100).

\section{Acknowledgments}
This research was kindly supported by the Quantum Computing Consortium that has been funded by the MAGNET program of the Israel Innovation Authority (IIA).

\bibliographystyle{eplbib}
\bibliography{references}

\begin{thebibliography}{10}
\expandafter\ifx\csname url\endcsname\relax\def\url#1{\texttt{#1}}\fi

\bibitem{HHL2009}
\Name{Harrow A.~W., Hassidim A. \and Lloyd S.} \REVIEW{Physical Review Letters}{103}{2009}{}.

\bibitem{Cao2012}
\Name{Cao Y., Anmer D., Frankel S.~H. \and Sabre K.} \REVIEW{Molecular Physics}{110}{2012}{1675}.

\bibitem{bharadwaj2023PNAS}
\Name{Bharadwaj S.~S. \and Sreenivasan K.~R.} \REVIEW{Proceedings of the National Academy of Sciences}{120}{2023}{e2311014120}.

\bibitem{ingelmann2024}
\Name{Ingelmann J., Bharadwaj S.~S., Pfeffer P., Sreenivasan K.~R. \and Schumacher J.} \REVIEW{Computers \& Fluids}{}{2024}{106369}.

\bibitem{bharadwaj2024compact}
\Name{Bharadwaj S.~S. \and Sreenivasan K.~R.} \REVIEW{arXiv preprint arXiv:2405.09767}{}{2024}{}.

\bibitem{kuznik2010lbm}
\Name{Kuznik F., Obrecht C., Rusaouen G. \and Roux J.-J.} \REVIEW{Computers \& Mathematics with Applications}{59}{2010}{2380}.

\bibitem{Steijl2022}
\Name{Steijl R.} \REVIEW{Applied Sciences 2023, Vol. 13, Page 529}{13}{2022}{529}.
\newline\url{https://www.mdpi.com/2076-3417/13/1/529/htm https://www.mdpi.com/2076-3417/13/1/529}

\bibitem{steijl2024floating}
\Name{Steijl R.} \REVIEW{}{}{2024}{}.

\bibitem{Mezzacapo2015}
\Name{Mezzacapo A., Sanz M., Lamata L., Egusquiza I.~L., Succi S. \and Solano E.} \REVIEW{Scientific Reports 2015 5:1}{5}{2015}{1}.
\newline\url{https://www.nature.com/articles/srep13153}

\bibitem{budinski2021ADE}
\Name{Budinski L.} \REVIEW{Quantum Information Processing}{20}{2021}{57}.

\bibitem{budinski2022NSE}
\Name{Budinski L.} \REVIEW{Int. J. Quantum Information}{20}{2022}{2150039}.

\bibitem{Schalkers2024JCP}
\Name{Schalkers M.~A. \and Möller M.} \REVIEW{J. Comp. Phys.}{502}{2024}{112816}.

\bibitem{dinesh2024_linear}
\Name{Dinesh~Kumar E. \and Frankel S.~H.} \REVIEW{arXiv:2405.08669}{}{2024}{}.

\bibitem{Succi2023}
\Name{Succi S., Itani W., Sreenivasan K. \and Steijl R.} \REVIEW{Europhysics Letters}{144}{2023}{10001}.
\newline\url{https://dx.doi.org/10.1209/0295-5075/acfdc7}

\bibitem{carleman1932}
\Name{Carleman T.} \REVIEW{}{}{1932}{}.

\bibitem{itani2022}
\Name{Itani W. \and Succi S.} \REVIEW{Fluids}{7}{2022}{24}.

\bibitem{itani2024}
\Name{Itani W., Sreenivasan K.~R. \and Succi S.} \REVIEW{Phys. Fluids}{36}{2024}{017112}.
\newline\url{https://doi.org/10.1063/5.0176569}

\bibitem{Claudio2024POF}
\Name{Sanavio C., Scatamacchia R., de~Falco C. \and Succi S.} \REVIEW{Physics of Fluids}{36}{2024}{}.

\bibitem{Claudio2024}
\Name{Sanavio C. \and Succi S.} \REVIEW{AVS Quantum Science}{6}{2024}{023802}.
\newline\url{https://doi.org/10.1116/5.0195549}

\bibitem{Toro2013}
\Name{Toro E.~F.} \Book{Riemann solvers and numerical methods for fluid dynamics: a practical introduction} (Springer Science \& Business Media) 2013.

\bibitem{shende2005}
\Name{Shende V.~V., Bullock S.~S. \and Markov I.~L.} \Book{Synthesis of quantum logic circuits} in proc. of \Book{Proceedings of the 2005 Asia and South Pacific Design Automation Conference} 2005 pp. 272--275.

\end{thebibliography}

\end{document}